\documentstyle[preprint,  aps, eqsecnum, color]{revtex}

\tightenlines

\definecolor{Red}{rgb}{1.00, 0.00, 0.00}
\definecolor{Green}{rgb}{0.00, 1.00, 0.00}
\definecolor{Blue}{rgb}{0.00, 0.00, 1.00}
\definecolor{Cyan}{rgb}{0.00, 1.00, 1.00}
\definecolor{Magenta}{rgb}{1.00, 0.00, 1.00}
\definecolor{Yellow}{rgb}{1.00, 1.00, 0.00}
\definecolor{DarkRed}{rgb}{0.7, 0.0, 0.0}
\definecolor{DarkBlue}{rgb}{0.00, 0.00, 0.70}
\definecolor{DeepSkyBlue}{rgb}{0.00, 0.75, 1.00}
\definecolor{DarkGreen}{rgb}{0.00, 0.50, 0.00}
\definecolor{SpringGreen}{rgb}{0.00, 0.80, 0.50}
\definecolor{Orange}{rgb}{1.00, 0.5, 0.00}
\definecolor{DarkOrange}{rgb}{0.8, 0.4, 0.00} 
\definecolor{OrangeRed}{rgb}{1.00, 0.5, 0.00} 
\definecolor{DeepPink}{rgb}{1.00, 0.08, 0.57}
\definecolor{DarkViolet}{rgb}{0.5, 0.00, 0.6}
\definecolor{SaddleBrown}{rgb}{0.54, 0.27, 0.07}
\definecolor{Gold}{rgb}{1.00, 0.84, 0.00} 

\begin{document}

%
%
%
%
\def\wbar{\bar{w}}
\def\Langle{\left\langle}
\def\Rangle{\right\rangle}
\def\oti{{\otimes}}
\def\bra#1{{\langle #1 |  }}
\def\lb{ \left[ }
\def\rb{ \right]  }
\def\tilde{\widetilde}
\def\bar{\overline}
\def\hat{\widehat}
\def\*{\star}
\def\[{\left[}
\def\]{\right]}
\def\({\left(}		\def\BL{\Bigr(}
\def\){\right)}		\def\BR{\Bigr)}
	\def\BBL{\lb}
	\def\BBR{\rb}
%
%
\def\zb{{\bar{z} }}
\def\zbar{{\bar{z} }}
\def\frac#1#2{{#1 \over #2}}
\def\inv#1{{1 \over #1}}
\def\half{{1 \over 2}}
\def\d{\partial}
\def\der#1{{\partial \over \partial #1}}
\def\dd#1#2{{\partial #1 \over \partial #2}}
\def\vev#1{\langle #1 \rangle}
\def\ket#1{ | #1 \rangle}
\def\rvac{\hbox{$\vert 0\rangle$}}
\def\lvac{\hbox{$\langle 0 \vert $}}
\def\2pi{\hbox{$2\pi i$}}
\def\e#1{{\rm e}^{^{\textstyle #1}}}
\def\grad#1{\,\nabla\!_{{#1}}\,}
\def\dsl{\raise.15ex\hbox{/}\kern-.57em\partial}
\def\Dsl{\,\raise.15ex\hbox{/}\mkern-.13.5mu D}
%
%
\def\th{\theta}		\def\Th{\Theta}
\def\ga{\gamma}		\def\Ga{\Gamma}
\def\be{\beta}
\def\al{\alpha}
\def\ep{\epsilon}
\def\vep{\varepsilon}
\def\la{\lambda}	\def\La{\Lambda}
\def\de{\delta}		\def\De{\Delta}
\def\om{\omega}		\def\Om{\Omega}
\def\sig{\sigma}	\def\Sig{\Sigma}
\def\vphi{\varphi}
%
%
\def\CA{{\cal A}}	\def\CB{{\cal B}}	\def\CC{{\cal C}}
\def\CD{{\cal D}}	\def\CE{{\cal E}}	\def\CF{{\cal F}}
\def\CG{{\cal G}}	\def\CH{{\cal H}}	\def\CI{{\cal J}}
\def\CJ{{\cal J}}	\def\CK{{\cal K}}	\def\CL{{\cal L}}
\def\CM{{\cal M}}	\def\CN{{\cal N}}	\def\CO{{\cal O}}
\def\CP{{\cal P}}	\def\CQ{{\cal Q}}	\def\CR{{\cal R}}
\def\CS{{\cal S}}	\def\CT{{\cal T}}	\def\CU{{\cal U}}
\def\CV{{\cal V}}	\def\CW{{\cal W}}	\def\CX{{\cal X}}
\def\CY{{\cal Y}}	\def\CZ{{\cal Z}}

\def\rvac{\hbox{$\vert 0\rangle$}}
\def\lvac{\hbox{$\langle 0 \vert $}}
\def\comm#1#2{ \BBL\ #1\ ,\ #2 \BBR }
\def\2pi{\hbox{$2\pi i$}}
\def\e#1{{\rm e}^{^{\textstyle #1}}}
\def\grad#1{\,\nabla\!_{{#1}}\,}
\def\dsl{\raise.15ex\hbox{/}\kern-.57em\partial}
\def\Dsl{\,\raise.15ex\hbox{/}\mkern-.13.5mu D}
%
%
%
\font\numbers=cmss12
\font\upright=cmu10 scaled\magstep1
\def\stroke{\vrule height8pt width0.4pt depth-0.1pt}
\def\topfleck{\vrule height8pt width0.5pt depth-5.9pt}
\def\botfleck{\vrule height2pt width0.5pt depth0.1pt}
\def\Zmath{\vcenter{\hbox{\numbers\rlap{\rlap{Z}\kern
0.8pt\topfleck}\kern 2.2pt
                   \rlap Z\kern 6pt\botfleck\kern 1pt}}}
\def\Qmath{\vcenter{\hbox{\upright\rlap{\rlap{Q}\kern
                   3.8pt\stroke}\phantom{Q}}}}
\def\Nmath{\vcenter{\hbox{\upright\rlap{I}\kern 1.7pt N}}}
\def\Cmath{\vcenter{\hbox{\upright\rlap{\rlap{C}\kern
                   3.8pt\stroke}\phantom{C}}}}
\def\Rmath{\vcenter{\hbox{\upright\rlap{I}\kern 1.7pt R}}}
\def\Z{\ifmmode\Zmath\else$\Zmath$\fi}
\def\Q{\ifmmode\Qmath\else$\Qmath$\fi}
\def\N{\ifmmode\Nmath\else$\Nmath$\fi}
\def\C{\ifmmode\Cmath\else$\Cmath$\fi}
\def\R{\ifmmode\Rmath\else$\Rmath$\fi}







\def\flux{\Phi}
\def\vep{\varepsilon}

\title{\color{DarkViolet} 
Gauge Invariance and the Critical Properties of 
Quantum Hall Plateaux Transitions} 
\author{\color{DarkGreen} Andr\'e  LeClair}
\address{Newman Laboratory, 
Cornell University, Ithaca, NY
14853.}
\date{\today}
\maketitle

\begin{abstract}

{\color{DarkBlue} 
A  model consisting of a single massless scalar field with
a topological coupling to a pure gauge field is defined 
and studied.  It possesses an $SL(2,\Zmath)$ symmetry as 
a consequence of the gauge invariance.  We propose that by
adding impurities the   
model 
can be used to  describe transitions between Quantum
Hall plateaux.  This leads to a correlation length exponent of} 
{\color{DarkRed} $20/9$},  {\color{DarkBlue} 
in excellent agreement with the most recent experimental
measurements. }  

\end{abstract}

\vskip 0.2cm
\narrowtext

\def\DG{DarkGreen}

{\color{DarkGreen} \section{ Introduction} }

Theoretical studies of the  Quantum Hall
 effect can roughly be classified
either as studies of the many-body physics on a plateau (Laughlin
wave function, etc.) or as studies of the plateau to plateau 
transitions.  For a review of the latter see \cite{Huckestein}. 
The transitions have been viewed as metal-insulator quantum phase
transitions and viewed as such are  problems in Anderson localization
\cite{Anderson0}\cite{Anderson}.  
The theoretical framework developed to study these
problems involves the study of electrons in a disordered potential.
Averaging over disorder using replicas  leads to a sigma model 
in 2 dimensions
\cite{Wegner}.

In applying this framework to the Quantum Hall transitions important
progress was made by Pruisken \cite{Pruisken}\cite{Prange}
 who understood that
a topological term proportional to the Hall conductivity $\sigma_{xy}$ 
was essential for obtaining the correct features of the phase diagram.
For the supersymmetric version see \cite{Zirnbauer}. 
The topological term modifies significantly the renormalization group
and hence the infra-red properties, in the same way that the 
$O(3)$ non-linear sigma model with topological coupling at $\theta=\pi$ 
has a non-trivial infra-red fixed point\cite{Haldane}.

Despite this progress, the detailed critical properties, such as the
exponents, have so far remained uncomputable.  This is due to the
complexity of the sigma models obtained and the usual difficulty
in finding non-trivial fixed points in the infra-red.  It remains
unclear whether the correct critical exponents are even contained
in the sigma-models.  

Progress has been made by inventing a network model which is believed to
be in the same universality class as the transition\cite{Chalker}.  
Here the transition resembles percolation through the impurities.  
The network model
was recently shown\cite{Chalker2} to be equivalent to fermions
with Dirac-like hamiltonian with various random potentials of the kind
studied in\cite{Ludwig}.  For these models  one is again  unable to 
compute the critical exponents analytically
 due to the complexities introduced
by disorder averaging and one has to resort to numerical 
methods.

On the other hand, 
very general and elegant  arguments, based largely on
gauge invariance,  were given 
 by  Laughlin and Halperin  to explain
the very existence of the plateaux\cite{Laughlin}\cite{Halperin}.
The argument does not depend on any detailed properties of
the  disorder;  it just has to be there to localize some states.  
Indeed, Halperin argued that 
 from the gauge invariance alone, one can infer the existence
of both localized and extended states;  the argument does not involve
finding a critical strength of the disorder for a localization/delocalization
phase transition.    
What helps the situation  is the fact that  in the    scaling approach
to Anderson localization, two dimensions is distinguished. 
In principle,  when there is no magnetic field,
 states in two dimensions are always localized, no matter
what the strength of the disorder\cite{Fisher}.  
The most extreme conclusion to be reached based on these observations
is that perhaps even the critical exponents governing the transitions
between plateaux may follow largely from gauge invariance. It would
indeed be very satisfying if both the existence of the plateaux 
and the critical properties of the transitions between plateaux 
were consequences of the same fundamental principle of gauge invariance. 
In 
this paper I  will construct a model based on this notion.    
In other words, we will simply assume disorder is irrelevant. 
The issue of whether disorder is relevant or irrelevant in
the renormalization group  will be
addressed in a separate publication\cite{andre}.

In the approach developed here,  in a sense I start from the  
end (the critical theory) rather than from the beginning.  
Namely, we construct a conformal  theory that contains the essential
features of the extended states.  We then add impurities as
a perturbation.  What is normally considered in the literature
is the reverse, i.e. one starts from the localized states and looks
for a delocalization transition.  
Critical theories in two dimensions are strongly constrained by 
conformal invariance\cite{BPZ}, and the computation of the critical
exponents should amount to the proper exercise in conformal field
theory.  This involves identifying the appropriate class of conformal
field theory.  
Guided by the importance of gauge invariance,  
the significance of topological terms (the topological
term in the sigma-model approach should survive in the infra-red
since it is proportional to $\sigma_{xy}$), 
we  construct a certain model of conformal field theory   
which possesses the necessary ingredients.  The hamiltonian 
corresponds to 2-component Dirac fermions.   The definition of the model
includes  a constraint on the zero modes that incorporates
an essential feature of the
 $2+1$ dimensional Quantum Hall dynamics.  We view this model
as describing the essential features of the extended states in
a {\it pure}  Quantum Hall system. 
   After bosonizing
the fermions and taking the gauge field to be a pure,  but singular,
gauge, we obtain a model with a topological coupling 
 similar to the multi-boson models studied originally in the context of
string theory\cite{Witten}\cite{Wilczek}.  However there are some
important differences.  Whereas the simplest (smallest) model studied
in \cite{Witten}\cite{Wilczek} has Virasoro central charge $c=2$
(two scalar fields),  the gauge invariance of our model effectively reduces
the degrees of freedom to a $c=1$ theory.  This is a new class of
conformal field theory
 characterized by an electric/magnetic  $SL(2,\Zmath)$ duality of the kind 
found by Cardy\cite{Cardy1} which acts 
on the
modular parameter $\tau = \theta/2\pi + i g^2 /2$, where $g,\theta$ 
are couplings in our model.  
A number of works have postulated a role for $SL(2,\Zmath)$ in
the Quantum Hall transitions, based on the phenomenology of 
transition selection rules\cite{Kiv1}\cite{dual}\cite{Frad}
\cite{dual4}\cite{dual5}. 
Though our microscopic 
model does not
have some  of the features assumed in these works, it may perhaps
still be useful towards developing these ideas. 
Our model is easily studied without utilizing the
full machinery of conformal field theory;  the elementary tools we need 
are bosonization\cite{Coleman} and the properties of exponentials
of massless scalar fields, reviewed in \cite{Ginsparg}\footnote{For
a more comprehensive review see\cite{Mathieu}.}.

Starting from the Kubo formula in $2+1$ dimensions we   derive
some formulas for the conductivities $\sigma_{xx}$, $\sigma_{xy}$ 
which involve current-current correlation functions  in the 2 dimensional 
quantum field theory.  These formulas are different from the kind
of formulas 
found in the literature, the latter being expressed 
as a double
product of  retarded
and advanced 1-particle Green functions. We argue that  our
formulas  represent
the conformal contribution to the conductivity.  
In our model the conductivities are functions
of $g,\theta$, and when $\sigma_{xx} =0$, 
$\theta = 1/\sigma_{xy}$.

We then add an impurity potential
 as a perturbation away from the critical point. 
If, as is normally done, one takes a gaussian random potential,
then according to the usual approach to   Anderson localization,
one must average over the disorder and search for a critical point
corresponding to the critical strength of the disorder for a
localization/delocalization phase transition.  
For the reasons given above, we chose the potential so that no
disorder averaging is needed.  
In  real experiments the critical exponents
are measured for a single sample, 
whereas disorder averaging amounts to averaging over
different realizations of disorder.  
If we assume that disorder is irrelevant, then 
universality implies
that {\it any} impurity potential which breaks translational invariance
should suffice.  Hence, viewing the sample as a very large disk, we
take the impurity potential to be a circle of impurities somewhere
inside the disk.  This  Corbino disk geometry has essentially all
of the ingredients used   in the gauge
argument of  Halperin\cite{Halperin}. 
The virtue of doing this is that the critical properties do not
depend on the strength of the impurity potential. 
Furthermore we are already at a fixed point 
before perturbation.  
 Our resulting model can be solved by mapping it
onto a boundary field theory, with the impurities residing at the 
boundary.  

\def\beg{\begin{equation}}
\def\endeq{\end{equation}}

In the scaling theory developed in\cite{Pruisken2}, the primary
critical exponents are $\mu, \nu$ defined as 
\beg
\label{1.1}
\Delta B \propto T^{\mu}, ~~~~~
\( \frac{\d \rho_{xy} }{\d B} \)_{\rm max} \propto T^{-\mu} 
\endeq
and also 
\beg
\label{1.2}
\xi_c \propto |B-B_c|^{-\nu} 
\endeq
Here, $\rho_{xy}$ is the resistivity, $B$ the magnetic field, 
$\Delta B$ the width of the region between plateaux, $B_c$ 
the  critical value at a transition,  $T$ the
temperature and $\xi_c$ the correlation length. The exponents are
related by $\mu = p/2\nu$, where $p$ is the inelastic scattering
length exponent.  In the limit of zero temperature, one expects $\mu = 1/\nu$.
 It has been shown
in experiments that the exponents  are
 independent of which plateaux are involved,
including the fractional ones.  As we will see, our model has this feature.  
We will also argue that our model predicts the value:  
\beg
\label{1.3}
\nu = 20/9 
\endeq

There has been a certain prejudice for the value $\nu = 7/3 = 21/9$, 
though the only analytical  computation supporting this value\cite{fourseven}
is based on the  percolation picture of the network model\footnote{$\nu
= 7/3 
= \nu_p + 1$, where $\nu_p = 4/3$ is the percolation exponent.}, 
and furthermore 
has been criticized as perhaps not corresponding
to the right physics\cite{Huckestein}.  Experimental measurements of
$\nu = 2.4 \pm 0.1$ and $\nu = 2.3 \pm 0.1$ were reported in \cite{Wei}
and \cite{Koch} respectively.
If we are  
 allowed to assume $\mu=1/\nu $, then our result is rather close
the most recent measurement of $\mu = .45 \pm .05$ 
reported in \cite{Sondhi}. 
The numerical estimate with the smallest error, $\nu = 2.35 \pm .03$ 
is due to Huckestein and includes data from both random Landau matrix
approach to an Anderson tight-binding model in a magnetic field 
and from the network model\cite{Huckestein}.\footnote{The data from the 
random  Landau
matrices and from the network model were statistically indistinguishable.
(B. Huckestein, private communication.)}  
A variant of the network model\cite{Kivelson}
gives $\nu = 2.43\pm .18$.  Our value is closer to the result of Ando
$\nu = 2.2 \pm .1$ obtained from transmission through a disordered
system\cite{Ando}.    
Indeed if the result (\ref{1.3}) is
actually correct then the consistently higher  values for $\nu$
obtained numerically
for the network/Anderson models
suggests that our model is in a different universality
class.   Certainly it is clear from the very 
definitions of the models that they are not simply equivalent.

\vfill\eject

{\color{\DG} 
\section{ The critical model and its conformal properties}
}

\def\DV{DarkViolet}

{\color{\DV} 
\subsection{ Definition of the Model}
}

\def\d{\partial}

We consider fermions in two spacial dimensions with coordinates
$x,y$, which we often denote simply as $x$.  
The constants $e=c=\hbar = 1$.  The second quantized action
one needs to study is 
\begin{equation}
\label{2.1}
S_{2+1} = \int dt d^2 x ~ \Psi^\dagger \( i \d_t - H \) \Psi 
\end{equation}
where $H$ is the hamiltonian.   For the purpose of
studying disorder in the $x-y$ plane arising from a disordered potential
in $H$, it is convenient to work with the time-Fourier transformed
Green functions.  Thus we expand
\begin{equation}
\label{2.2} 
\Psi(x,t) = \int  \frac{d\vep}{\sqrt{2\pi}} ~ e^{-i \vep t} \, \Psi_\vep (x) 
\end{equation}
and the action becomes 
\begin{equation}
\label{2.3} 
S_{2+1} = \int d\vep \int d^2 x ~ \Psi^\dagger_\vep (x) (\vep - H ) 
\Psi_\vep (x)
\end{equation} 
Since the functional integral is defined by $e^{iS}$, for a fixed energy
$\vep$ one needs to study a euclidean functional integral defined by
$e^{-S}$, where 
\begin{equation}
\label{2.4} 
S = i \int d^2x ~ \Psi^\dagger (H-\vep) \Psi  
\end{equation}
and  we have dropped the subscript $\vep$ on $\Psi$.  

\def\beg{\begin{equation}}
\def\endeq{\end{equation}}
\def\dx{{ \frac{d^2x}{2\pi} }}

Our model consists of a two component fermion 
\begin{equation}
\label{2.4b}
\Psi = \left( \matrix{\psi_1 \cr \psi_2\cr } \right) 
\end{equation}
with the hermitian Dirac hamiltonian
\begin{equation}
\label{2.5}
H = \inv{\sqrt{2}} \( -i \d_x - A_x \) \sigma_x 
+ \inv{\sqrt{2}} \( -i \d_y - A_y \) \sigma_y + V(x,y) 
\end{equation}
This kind of Dirac hamiltonian has been considered before in connection
with Quantum Hall transitions\cite{Ludwig}\cite{Chalker2}, but here
the meaning is rather different.  Our  theory is interpreted
as describing the extended states in a bulk
that is free of impurities.  For spinless electrons it should be 
a scale invariant theory with $c=1$ that is rotationally invariant,
and the above hamiltonian is essentially the unique one with these 
properties.   Wen's description of the edge states may be useful
in making this connection more precise\cite{Wen}.  
Defining  the complex coordinates
\begin{equation}
\label{2.6}
z = \inv{\sqrt{2}} (x+ i y) , ~~~~~ \zbar = \inv{\sqrt{2}} (x-iy),  
\endeq
and the gauge fields $A_z = (A_x - i A_y )/\sqrt{2}$, 
$A_\zbar = (A_x + i A_y )/\sqrt{2}$, upon  
rescaling $\Psi \to \Psi /\sqrt{2\pi}$ one obtains 
\beg
\label{2.7} 
S = \int \dx \[ \psi_1^\dagger (\d_z -i A_z ) \psi_2 
+ \psi_2^\dagger (\d_\zbar - i A_\zbar ) \psi_1 
- i (V+\vep) \( \psi_2^\dagger \psi_2 + \psi_1^\dagger \psi_1 \) \] 
\endeq
The $V$ and $\vep$ terms give the fermions a mass and thus break
the conformal invariance.  For the remainder of this section we
set $V=\vep=0$ and focus on the conformal field theory coupled
to the gauge field.  

The fermions can be bosonized with a single scalar field $\phi$
satisfying $\d_z \d_\zbar \phi = 0$.  The scalar field separates 
into left and right moving parts
\beg
\label{lr}
\phi (z, \zbar) = \phi_L (z) + \phi_R (\zbar) 
\endeq
and the fermions have the bosonized expressions
\beg
\label{2.8} 
\psi_1 = e^{-i\phi_L}, ~~~~~ \psi_2^\dagger = e^{i\phi_L }, ~~~~~ 
\psi_2 = e^{i\phi_R}, ~~~~~ \psi_1^\dagger = e^{-i\phi_R } 
\endeq
The current coupled to the gauge field then has the following 
bosonized expressions
\begin{eqnarray}
\label{2.9} 
j_z &=& \inv{2\pi} \psi_2^\dagger \psi_1 = ~\frac{i}{2\pi} \d_z \phi 
\\ \nonumber
j_\zbar &=& \inv{2\pi} \psi_1^\dagger \psi_2 = - \frac{i}{2\pi} \d_\zbar \phi
\end{eqnarray} 

Introducing the completely anti-symmetric tensor 
\beg
\label{anti}
\ep_{xy} = -\ep_{yx} ; ~~~~~\ep_{z\zbar} = - \ep_{\zbar z} = i 
\endeq
the current takes the form 
\beg
\label{2.10}
j_\mu = \inv{2\pi} \ep_{\mu\nu} \d_\nu \phi 
\endeq
The current is thus a topological current which is identically
conserved $\d_\mu j_\mu = 0$ by virtue of the anti-symmetry of 
$\ep_{\mu\nu}$.  
The bosonized action takes the form 
\beg
\label{2.11}
S = \int d^2 x ~ \[  \inv{8\pi} \d_\mu \phi \d_\mu \phi - \frac{i}{2\pi} 
\ep_{\mu\nu} \d_\nu \phi A_\mu \]  
\endeq
where 
$\d_\mu \d_\mu = \d_z \d_\zbar + \d_\zbar \d_z $. 

In order to clarify some of our subsequent arguments, we introduce
a coupling $g$ by rescaling the current in the $j_\mu A_\mu$ coupling
to 
\beg
\label{2.12}
j_\mu = \inv{2\pi g} \ep_{\mu\nu} \d_\nu \phi
\endeq
One justification for this is to consider adding 
a current-current interaction 
\beg
\label{int}
 \frac{\pi k}{2} \int d^2 x \, j_\mu j_\mu  = 
k \int d^2 x \, \inv{8\pi} (\d_\mu \phi )^2   
\endeq
Such an interaction would arise for instance upon averaging over
a gaussian disordered component of the gauge field $A_\mu$, where
$k$ is proportional to the variance. 
This interaction merely re-scales 
the kinetic term.  
Redefining $\phi \to \phi / \sqrt{1+k}$ leads to the current 
(\ref{2.12}) with $g^2 = 1+k$.  
If one prefers, $g$ can be viewed  as a kind of bookkeeping device
that will be convenient in the sequel.  
Our original model
corresponds to $g=1$.  


\def\phid{{\tilde{\phi}}}

Whereas the fermionic description has manifest gauge symmetry,
the bosonic description as it stands does not.  
However, the gauge invariance can be restored by adding an $A^2$ term:
\beg
\label{2.13}
S = \int d^2 x \[ \inv{8\pi} 
\d_\mu \phi \d_\mu \phi - \frac{i}{2\pi g} \ep_{\mu \nu} 
\d_\nu \phi A_\mu + \inv{2\pi g^2} A_\mu A_\mu \] 
\endeq
To see this, introduce the dual field $\phid$ defined as 
\beg
\label{2.14}
\d_\mu \phid = -i \ep_{\mu \nu} \d_\nu \phi 
\end{equation}
In terms of left and right movers defined in Eq. (\ref{lr}), the
above equation implies (up to a constant)
\beg
\label{2.14b}
\phid = \phi_L - \phi_R
\endeq
Using the relation $(\d_\mu \phi)^2 = - (\d_\mu \phid )^2 $ one
can verify that 
the action has the local gauge invariance
\beg
\label{2.15} 
\phid (x) \to \phid (x) + \frac{2}{g} \lambda (x) , 
~~~~~ A_\mu (x) \to A_\mu (x) + \d_\mu \lambda (x) 
\endeq
Since $\psi_{1} = \exp(-i (\phi + \phid)/2 )$, 
$\psi_2 = \exp (i (\phi - \phid )/2 ) $, the gauge transformation
on the fermions is 
\beg
\label{gaugeb}
\psi_{1,2} \to e^{-i\lambda} \psi_{1,2} 
\endeq
for $g=1$, as it should be.  The addition of the $A^2$ term is
analogous to the Green-Schwarz mechanism for anomaly cancelation in
low energy effective string theory\cite{Schwarz}.  

For application to the Quantum Hall effect, the gauge field
should incorporate some magnetic flux $\Phi$ through the
$x-y$ plane.  The flux is 
$\Phi = \int d^2 x \vec{\nabla} \times \vec{A} = \oint dx_\mu A_\mu$. 
A non-zero flux can be obtained with $A_\mu$ a pure gauge, $A_\mu 
= \d_\mu \chi$, as long as $\chi$ is allowed to be discontinuous 
across a cut in the $x-y$ plane extending from the origin to
infinity.  Namely,  the flux is non-zero if $\chi$ has winding
modes  
\beg
\label{Flux}
\Phi =  \int d\sigma \d_\sigma \chi = \chi(\sigma = 2\pi) - \chi(0), 
~~~~~  z = \frac{r}{\sqrt{2}} e^{i\sigma} 
\endeq
For $A_\mu =\d_\mu \chi$ the action takes the form 
\beg
\label{2.16}
S = \int d^2 x \[ \inv{8\pi} 
\d_\mu \phi \d_\mu \phi - \frac{i}{2\pi g} \ep_{\mu \nu} 
\d_\nu \phi \d_\mu \chi  + \inv{2\pi g^2}  \d_\mu \chi \d_\mu \chi \] 
\endeq

The coupling between $\chi$ and $\phi$ is a topological term, i.e.
a total derivative.  A model of two independent scalar fields 
with the same topological coupling has been studied before
as a $c=2$ conformal field theory\cite{Witten}\cite{Wilczek}.   
Our model differs in a significant way:  due to the structure of
the couplings the gauge invariance Eq. (\ref{2.15}) allows $\chi$ 
to be gauged away up to the effects of its discontinuity across 
the cut.  Thus we expect our model to share some of the features
found in \cite{Witten}\cite{Wilczek}, but in a $c=1$ conformal field
theory. 

As we now describe, a further constraint on the zero modes of $\chi$ 
allows us to incorporate an essential feature of the $2+1$ dimensional
Quantum Hall dynamics.  Electric current conservation implies the 
continuity equation $\d_t \rho + \d_\mu J_\mu = 0$ where 
$\rho$ is the charge density.  
When the conductivity $\sigma_{xx} = 0$ we may write 
$J_\mu = \sigma_{xy} \ep_{\mu\nu} E_\nu$.  Inserting this into the
continuity equation and using the Maxwell equation 
$\vec{\nabla} \times \vec{E} = - \d_t \vec{B}$, one obtains
$\d_t (\rho - \sigma_{xy} B ) = 0$, where $B$ is the magnetic
field perpendicular to the $x-y$ plane.  Integrating over space:
\beg
\label{2.16b}
\frac{d}{dt} \( q - \sigma_{xy} \Phi \) = 0
\endeq
where $q= \int d^2 x \rho $ is the electric charge and 
$\Phi = \int d^2 x B $ the flux.  

Let $\CC$ denote a circular contour of arbitrary radius surrounding
the origin.  If the charge inside this circle is zero at $t=-\infty$,
then at time $t$,
\beg
q(t) = - \int_{-\infty}^t dt \oint_\CC dx_\mu \ep_{\mu\nu} J_\nu  
\endeq
For $t \to \infty$, using the time dependence in Eq. (\ref{2.2}),
\beg
\int_{-\infty}^\infty dt J_\mu (x,t) = \int d\vep j_\mu (x)
\endeq
where $j_\mu$ is the current in Eq. (\ref{2.9}).  Since the flux
is $\Phi = \oint_\CC dx_\mu \d_\mu \chi$, for fixed $\vep$ the equation 
(\ref{2.16b}) leads to 
\beg
\label{2.17}
- \oint_\CC dx_\mu \ep_{\mu\nu} j_\nu = \sigma_{xy} 
\oint_\CC dx_\mu \d_\mu \chi 
\endeq

We take the point of view that the proportionality expressed in
Eq. (\ref{2.17}) captures an essential feature of the Quantum 
Hall dynamics.  Since this is real time dynamics in $2+1$
dimensions, it cannot follow from our action $S$, but must
be put in by hand.  We will express this proportionality
in terms of a fundamental parameter $\theta$: 
\beg
\label{2.18}
\frac{\theta}{2\pi g} \oint_\CC dx_\mu \d_\mu \phi = 
\oint_\CC dx_\mu \d_\mu \chi 
\endeq
where we have used Eq. (\ref{2.12}).  Though 
Eq. (\ref{2.18}) follows from Eq. (\ref{2.17}) with the identification
$\theta = 1/\sigma_{xy}$, we only expect this to be valid when
$\sigma_{xx} = 0$, and so it is incorrect to make this identification
at this stage.  Rather, in the next section we will compute the
conductivities $\sigma_{xx} , \sigma_{xy}$ in terms of the parameters
$g,\theta$. 

To summarize, our conformal model is defined by the action Eq. (\ref{2.16}) 
with the constraint Eq. (\ref{2.18}) on the zero modes.  

\vfill\eject

{\color{\DV} 
\subsection{ 
The Spectrum of Conformal Fields Depends on $\theta$}
}

The action possesses the gauge invariance
\beg
\label{gauge}
\chi \to \chi + \lambda, ~~~~~\phid \to \phid + \frac{2}{g} \lambda
\endeq
which follows from Eq. (\ref{2.15}).  Thus, $\chi$ can be gauged away
up to its effects on the zero modes.   
One consequence of this is that the spectrum of allowed fields is
modified in a way that depends on $\theta$.  Consider the 
conformal fields $\exp( i\alpha_L \phi_L + i \alpha_R \phi_R )$.  
The topological charge separates into left and right pieces:
\beg
\label{2.19}
Q = - \oint_\CC dx_\mu \ep_{\mu\nu} j_\nu = Q_L - Q_R 
\endeq
where 
\beg
\label{2.20}
Q_L = \inv{2\pi g} \oint dz  ~\d_z \phi_L , 
~~~~~
Q_R = \inv{2\pi g} \oint d\zbar ~ \d_\zbar \phi_R 
\endeq
The exponential fields are characterized by their charges
$\alpha_{L,R}$: 
\beg
\label{2.21} 
\[ Q_{L,R} , e^{i\alpha_L \phi_L + i \alpha_R \phi_R } \] 
= \alpha_{L,R} ~  
 e^{i\alpha_L \phi_L + i \alpha_R \phi_R }  
\endeq

To determine the spectrum of allowed $\alpha_{L,R}$ we conformally map the
theory onto the cylinder by letting 
\beg
\label{zcoor}
z = e^{w}, ~~~~~ w = t+ i\sigma 
\endeq
The anti-symmetric tensor is $\ep_{\sigma t} = -\ep_{t\sigma} = 1$.  
The coordinate $\sigma$ is along the circumference of the cylinder
and takes values $0\leq \sigma \leq 2\pi$, whereas $t$ runs along
the length of the cylinder, $-\infty \leq t \leq \infty$.  
In order to carry out canonical quantization, let us for the moment
further rotate to Minkowski space $t\to it$.  The field $\phi$ 
can be expanded as
\beg
\label{2.22}
\phi (t, \sigma) = \phi_0 (t) + g Q \sigma 
+ i \sum_n \inv{n} \( a_n e^{-in(t+\sigma)} + \bar{a}_n e^{-in(t-\sigma)} \)
\endeq
The lagrangian for the zero modes is 
\begin{eqnarray}
\label{2.23} 
L &=& \int_0^{2\pi} d\sigma \[ \inv{8\pi} 
\( \dot{\phi}^2_0 - g^2 Q^2 \) - \inv{2\pi g} \dot{\phi}_0 \d_\sigma \chi 
\]
\\ \nonumber
&=&    
\inv{4} \( \dot{\phi}_0^2 - g^2 Q^2 \) - \inv{2\pi g} \dot{\phi}_0 \Phi 
\end{eqnarray}
where $\Phi$ is again the magnetic flux.  

The momentum conjugate to $\phi_0$ is 
\beg
\label{2.24} 
p_0 = \inv{2} \dot{\phi}_0 - \inv{2\pi g} \Phi 
\endeq
Substituting $\phi_0 (t) = \phi_0 + \dot{\phi}_0 t$ into Eq. (\ref{2.22})
and using Eq. (\ref{2.24}) one obtains the zero mode contribution 
to $\phi$.  After mapping back to the $z$ coordinate one finds  
\beg
\label{2.25} 
\phi (z, \zbar) = \phi_0 -i Q_L \log z  - i Q_R \log \zbar + ...
\endeq
where 
\begin{eqnarray}
\label{2.25b} 
Q_L &=& \( p_0 + \frac{\Phi}{2\pi g} \) + \frac{gQ}{2} 
\\ \nonumber
Q_R &=& \( p_0 + \frac{\Phi}{2\pi g} \) - \frac{gQ}{2} 
\end{eqnarray}

We first impose that the charge $Q$ is an integer $n$.  Then, since
$[\phi_0 , p_0 ] = i$, mutual locality of the exponential fields
when $\theta = 0$ requires $p_0 = m/g$ where $m$ is some other integer. 
Finally the constraint Eq. (\ref{2.18}) imposes 
$\Phi = \theta Q$.  Thus the spectrum of exponential fields consists
of the operators 
\beg
\label{2.26}
\CO_{n,m} = \exp\( {i \alpha_L \phi_L + i \alpha_R \phi_R }\)
\endeq
where 
\begin{eqnarray}
\label{2.27} 
\alpha_L &=& \inv{g} \( m + \frac{\theta}{2\pi} n \) + \frac{gn}{2} 
\\ \nonumber
\alpha_R &=& \inv{g} \( m + \frac{\theta}{2\pi} n \) - \frac{gn}{2} 
\end{eqnarray}
The integers $n,m$ are electric and magnetic charges. 

The structure of the fields $\CO_{n,m}$ can be understood as simply
arising from the gauge transformation Eq. (\ref{2.15}).  Namely,
using Eq. (\ref{2.18}) to identify $\chi = \theta \phi/2\pi g$,
the gauge transformation reads
\beg
\label{2.28} 
\phid \to \phid  + \frac{\theta}{\pi g^2} \, \phi 
,~~~~~~~\phi \to \phi
\endeq
Let us  express
\beg
\label{2.30}
\alpha_L \phi_L + \alpha_R \phi_R = \alpha \phi + \tilde{\alpha} \phid
\endeq
where $\alpha = (\alpha_L + \alpha_R)/2$, $\tilde{\alpha} =
(\alpha_L - \alpha_R)/2$.  Then, indeed  one can verify that the
$\theta$ dependence in Eq. (\ref{2.27}) 
 follows from the shift Eq. (\ref{2.28}):
\beg
\label{2.31}
\CO_{n,m} (\phi, \phid; \theta) 
= \CO_{n,m} \(  \phi, ~   \phid + \frac{\theta}{\pi g^2} 
\phi ; ~ \theta=0 \) 
\endeq

In summary, our model is a free massless scalar $\phi$  supplemented 
by the transformation Eq. (\ref{2.28}).  
Any correlation function involving the fields $\phi, \phid$ will be 
computed by first performing the transformation (\ref{2.28}) 
and then using the identifications (\ref{lr})(\ref{2.14b})
and  the two-point functions:
\beg
\label{corr}
\langle \phi_L (z) \phi_L (0) \rangle = -  \log z
, ~~~~~~ 
\langle \phi_R (\zbar) \phi_R (0) \rangle = -  \log \zbar 
\endeq
The identifications (\ref{lr})(\ref{2.14b}) together with the above
equation imply (when $\theta = 0$):
\begin{eqnarray}
\label{corr2}
\langle \phi (z,\zbar) \phi (0) \rangle &=& \langle
\phid (z, \zbar) \phid (0) \rangle = - \log( z\zbar) 
\\ \nonumber
\langle \phi (z, \zbar) \phid (0) \rangle &=& 
\langle \phid (z,\zbar) \phi (0) \rangle = - \log\( \frac{z}{\zbar} \)  
\end{eqnarray}
Some  $\theta$ dependent correlation functions will be
computed in the sequel.

\bigskip

 {\color{\DV}   
\subsection{ The Partition Function has an
 $SL(2,\Zmath)$ Symmetry} 
}

The partition function on the torus possesses an 
$SL(2,\Zmath)$ modular symmetry acting on the coupling
constants $g,\theta$.  Demonstrating this in our 
($c=1$) conformal field theory closely parallels 
the discussion in \cite{Cardy1}\cite{Wilczek} for
2-boson ($c=2$) theories with topological term.  

A torus is obtained by imposing periodic boundary conditions on
the cylinder described above.  Let the length of the cylinder
in the $t$ direction be $l$.  Since the hamiltonian on 
the cylinder is 
$H = L_0 + \bar{L}_0 - c/12$, where $L_0, \bar{L}_0$ are the zero 
modes of the Virasoro algebra, one has
\beg
\label{Z1}
Z = {\rm Tr}  ~  e^{-l (L_0 + \bar{L}_0 - c/12 )} 
\endeq
The trace is over the Virasoro highest weight representations
corresponding to the fields $\CO_{n,m}$.  The zero mode contribution
to $L_0 + \bar{L}_0$ corresponds to the (anomalous) conformal 
scaling dimension of
$\CO_{n,m}$.
Using
\beg
\label{expcor}
\langle e^{i\alpha_L \phi_L (z)} e^{-i\alpha_L \phi_L (0)} \rangle 
= z^{-\alpha_L^2} , 
~~~~~
\langle e^{i\alpha_R \phi_R (\zbar)} e^{-i\alpha_R \phi_R (0)} \rangle 
= \zbar^{-\alpha_R^2} 
\endeq
one finds 
\beg
\label{Z2}
d_{n,m} (g,\theta) = \inv{2} ( \alpha_L^2 + \alpha_R^2 ) 
= \inv{g^2} \( m+ \frac{\theta n}{2\pi}  \)^2 + \frac{g^2 n^2}{4} 
\endeq
The partition function is then 
\beg
\label{Z3}
Z(g,\theta) = \inv{|\eta|^2} \sum_{n,m} \exp \( -l d_{n,m} (g,\theta) \)
\endeq
where  the Dedekind $\eta$-function comes from the non-zero modes
$a_n, \bar{a}_n$.  (See \cite{Ginsparg}\cite{Mathieu}.)  

\def\thpi{\frac{\theta}{2\pi}}

$Z$ is obviously invariant under $\theta \to \theta + 2\pi$ since this
just shifts the integer $m$.  It is also easy to show that 
\beg
\label{Z4} 
d_{n,m} (g, \theta) = d_{m, -n} (g', \theta' ) 
\endeq
with 
\begin{eqnarray}
\label{Z5}
g'^2 &=&  \frac{g^2}{\( g^4/4  + (\theta/2\pi )^2  \)}  
\\ \nonumber
\theta' &=& - \frac{\theta} { \(   g^4/4  + (\theta/2\pi )^2 \) }  
\end{eqnarray}
Thus, $Z$ is also invariant under $(g,\theta) \to (g' , \theta')$. 

Introduce a modular parameter $\tau$ (not to be confused with 
the geometrical modular parameter of the torus 
$\tau_{\rm torus} = i l/2\pi$)
\beg
\label{modular}
\tau = \thpi + i \frac{g^2}{2} 
\endeq
Then the two above symmetries correspond to 
\beg
\label{Z7}
\CT: ~~~~\tau \to \tau + 1 , ~~~~~\CS:~~~~ \tau \to - 1/\tau 
\endeq
These two transformations generate the group $SL(2, \Zmath)$, whose
elements $\Gamma$ transform $\tau \to \Gamma (\tau)= (a\tau + b)/(c\tau
+ d)$ where $a,b,c,d$ are integers satisfying $ad-bc = 1$.  
The partition function has the full symmetry
$Z(\Gamma(\tau)) = Z(\tau)$.   

\bigskip\bigskip

{\color{\DG} 
\section{ Conformal Conductivity} 
}

The conductivity tensor $\sigma_{\mu\nu}$ has the following properties:
$\sigma_{xx} = \sigma_{yy}$, $\sigma_{xy} = - \sigma_{yx}$.  In terms
of the complex coordinate $z$, this implies
\beg
\label{3.1}
\sigma_{zz} = \sigma_{\zbar \zbar} = 0
\endeq
The non-zero components are
\beg
\label{3.2}
\sigma_{\zbar z} = \sigma_{xx} + i \sigma_{xy} , 
~~~~\sigma_{z\zbar} = \sigma_{xx} + i \sigma_{yx} = (\sigma_{\zbar z})^* 
\endeq
Thus it is natural to think of the conductivity as the single
complex parameter $\sigma_{\zbar z}$.  
We will also use the covariant description
\beg
\label{3.3}
\sigma_{xx} = \inv{2} \sigma_{\mu \mu} , ~~~~~ 
\sigma_{xy} = \inv{2} \ep_{\mu\nu} \sigma_{\mu\nu} 
\endeq

 The conductivity
is usually expressed in terms of retarded and advanced
one-particle Green functions, which can be computed from the
action (\ref{2.7}) by including a small positive (negative) imaginary
part to $\vep$ for the retarded (advanced) Green function.   Studying a
localization/delocalization phase transition amounts to finding
a renormalization group fixed point of the theory where the theory
is conformally invariant.  Our aim in this section is to 
understand the properties of the conductance at the critical point,
thus we  
set the potential $V$ to zero.  The $\vep$-terms also break 
conformal invariance.  In what follows we will derive some simple 
expressions representing conformal contributions to the conductivity
when $V= \vep = 0$.  In a sense these ``conformal conductivities'' 
represent the conductivity at a possible fixed point of a model.
Later in the paper we will restore an impurity 
potential as a perturbation.  

We start from the Kubo formula in $2+1$ dimensions.  Throughout
this section $t$ is the real time in the $2+1$ dimensional world.
We work at finite
temperature as a computational tool, taking the zero temperature limit
at the end. The AC conductivity is given by 
\beg
\label{3.4} 
\sigma_{\mu\nu} (\omega) = \frac{i}{\omega} \Pi_{\mu\nu} (\omega) 
\endeq
where $\Pi (\omega)$ is an analytic continuation to real time of
the euclidean $(E)$  Matsubara quantity:
\beg
\label{3.5}
\Pi_{\mu\nu} (\omega) = \Pi^{(E)}_{\mu\nu} (i\omega \to \omega + i \eta )
\endeq
where $\eta$ is small and positive, and 
\beg
\label{3.6} 
\Pi^{(E)}_{\mu\nu} (i\omega) = -
\int d^2 x \int_0^\beta d\tau e^{i\omega \tau} 
\langle J_\mu (x,\tau) J_\nu (0) \rangle 
\endeq
Here, $\tau = it$ is euclidean time, $\beta$ is the inverse temperature, 
and the correlation is at finite temperature. 

In the Matsubara formulation we work with the action 
\beg
\label{3.7}
S = \int_0^\beta d\tau \int d^2 x ~ \psi^\dagger (\d_\tau + H ) \psi 
\endeq
and expand the fields as follows:
\beg
\label{3.8}
\psi (x,\tau) = \sum_\nu e^{-i\nu \tau} \, \psi_{i\nu} (x) 
\endeq
where $\nu = 2\pi(n+ 1/2)/\beta$ with $n$ an integer.  The action
then takes the form 
\beg
\label{3.9} 
S = \sum_\nu \beta \int d^2 x  ~ \psi^\dagger_{i\nu} (x) 
 (H-i\nu) \psi_{i\nu} (x)
\endeq

For notational
 simplicity, we carry out the computation without displaying the
spacial tensorial properties, restoring them at the end.  The current
is a fermion bilinear:
\beg
\label{3.10} 
J(x,\tau) = \sum_{\nu, \nu'} e^{i(\nu'-\nu)\tau} ~ 
\psi^\dagger_{i\nu'} (x) \psi_{i\nu} (x) 
\endeq
Inserting this into Eq. (\ref{3.6}) one obtains 
\beg
\label{3.11} 
\Pi^{(E)} (i\omega) = i (e^{i\omega \beta} - 1 ) 
\sum_{\nu, \nu'} \int d^2 x \inv{\omega + \nu' - \nu } 
\Langle \psi^\dagger_{i\nu'} (x) \psi_{i\nu} (x) J(0) \Rangle 
\endeq

There are well-known techniques for carrying out the sums over
Matsubara frequencies (see e.g.  \cite{Mahan}).  The sum over $\nu'$ 
may be performed by considering the contour integral 
\beg
\label{3.12} 
\oint \frac{dz}{2\pi i}  ~ n_F (z) \,  
\inv{ z - (i\nu - i\omega)} ~ 
\Langle \psi^\dagger_z (x) \psi_{i\nu} (x) J(0) \Rangle 
\endeq
where
\beg
\label{3.13} 
n_F (z) = \inv{ e^{\beta z} + 1 } 
\endeq
and the contour of integration is a circle of radius $R$ as $R \to \infty$. 
In this way one picks up the poles in $n_F (z)$ at $z=i\pi (2n+1)/\beta$ 
with residue $-1/\beta$.  Assuming the only other pole is at 
$z= i\nu - i\omega$, one obtains
\beg
\label{3.14}
\Pi^{(E)} (i\omega ) = -\beta e^{i\omega \beta} 
\int d^2 x \sum_\nu 
\Langle \psi^\dagger_{i\nu - i\omega} (x) 
\psi_{i\nu} (x)  J (0) \Rangle
\endeq

The summation over $\nu$ is now studied by considering the contour
integral 
\beg
\label{3.15}
\oint \frac{dz}{2\pi i } ~ n_F (z) \, 
\Langle \psi^\dagger_{z-i\omega} (x) \psi_z (x) J(0) \Rangle 
\endeq
The integrand is known to have branch cuts at $z= \vep + i\omega$ and
$z= \vep$, where $\vep$ is real.  Thus the contour must be
chosen to run above and below the branch cuts along
$z= \vep + i\omega \pm i\delta$ and $z = \vep \pm i \delta$.  
The other parts of the contour are along a circle at $\infty$,
thus the integral Eq.(\ref{3.15}) is a sum over three closed 
contours and all poles in $n_F$ are picked up.
The sum over $\nu$ is then expressed as the sum of four real
integrations over $\vep$ coming from above and below the
branch cuts.  

What is normally done is to factorize the correlation function in 
Eq. (\ref{3.14}) into a product of two 1-particle Green functions;
the  result is an expression involving  a product of 
the difference between retarded and advanced Green functions, i.e.
the spectral density\footnote{For a discussion in the context of
disordered electrons, see \cite{Thouless}.}.  
The retarded (advanced) Green functions
correspond to taking $\eta = 0^+$ ( $0^-$)  in 
Eq. (\ref{3.5}).   This would lead one to believe that
when $\eta =0$ the conductivities are zero.  However as explained in
\cite{Stone} a non-zero density of states requires a vacuum
expectation value which survives when $\eta = 0$, and an analogy
has been made with spontaneous symmetry breaking\footnote{See
also section IVC.}.   
Our aim is to extract the conformal (critical)  contribution to
the conductivity, i.e. the part that survives as $\vep, \eta$ go
to zero, so we follow
a different procedure than the usual one. 
We are interested in the DC conductivity
at $\omega = 0$.  As $\omega$ tends to zero, in particular when 
$\omega = 2\delta$, the integrations along $z= \vep + i\omega - i\delta$
and $z= \vep + i\delta$ coalesce, and there is effectively only
one branch cut at $z=\vep$.  Integrations above and below this cut
give: 
\beg
\label{3.16}
\Pi^{(E)} (i\omega) = \beta^2 e^{i\omega \beta} \int d^2 x 
\int  \frac{d\vep}{2\pi i}
\( n_F(\vep + i\omega) - n_F (\vep) \) 
\Langle J_\vep (x) J_\vep (0) \Rangle 
\endeq
with $J_\vep = \psi^\dagger_\vep \psi_\vep $ where 
$\vep$ has the same meaning as in Eq. (\ref{2.3}).  
We can now make the analytic continuation 
(\ref{3.5}) and take the $\omega \to 0$ limit.  Using 
$\d_\vep n_F (\vep) = - \delta (\vep) $ at zero temperature, 
the integration over
$\vep$ sets $\vep =0$. Recall $\vep = 0$ corresponds to the
conformal limit.    Finally, rescaling 
$J_{\vep = 0} (x) \to j(x)/\beta$, the zero temperature limit may
be taken.  Our final result is the simple formula
\beg
\label{3.17} 
\sigma'_{\mu\nu} = - \int d^2 x ~ \Langle j_\mu (x) j_\nu (0) \Rangle 
\endeq
We denote the above quantity as $\sigma'_{\mu\nu}$ to emphasize
that it is the critical contribution to $\sigma_{\mu\nu}$. 
(The disappearance of the $1/2\pi $ in Eq. (\ref{3.16}) comes from the 
extra $2\pi$ in Eq. (\ref{2.7}).) 
The same rescaling of the currents by $1/\beta$ removes the $\beta$ 
in Eq. (\ref{3.9}).  Thus in the formula Eq. (\ref{3.17}), the currents
$j_\mu$ are those in Eq. (\ref{2.9}) and the correlation function 
is computed with respect to the action (\ref{2.16}) with $\vep = 0$.   
The conformal currents $j_\mu$ have dimension $1$, so that 
$\sigma'_{\mu\nu}$
is dimensionless as it should be.

We now evaluate $\sigma'_{\mu\nu}$ in our model.  The current is 
$j_\mu = i \d_\mu \phid/2\pi g$.  We evaluate the current-current
correlator in the presence of the coupling to $\chi$ by simply
performing the gauge transformation Eq. (\ref{2.28}): 
\begin{eqnarray}
\label{3.18}
\sigma'_{xx} &=& \inv{8\pi^2 g^2} \int d^2 x ~  
\Langle \d_\mu \( \phid(x) + \frac{\theta}{\pi g^2} \phi (x) \)  
\d_\mu \( \phid (0) + \frac{\theta}{\pi g^2} \phi (0) \) \Rangle 
\\ \nonumber
\sigma'_{xy} &=& \inv{8\pi^2 g^2} \int d^2 x ~ \ep_{\mu\nu} \,  
\Langle \d_\mu \( \phid(x) + \frac{\theta}{\pi g^2} \phi (x) \)  
\d_\nu \( \phid (0) + \frac{\theta}{\pi g^2} \phi (0) \) \Rangle 
\end{eqnarray}

We take the geometry to be a disk of radius $r$ with $r$ going
to $\infty$, and   we don't impose any specific boundary condition
at infinity.    The integrals in Eq. (\ref{3.18}) can be moved to
the boundary of the disk using the divergence and 
Stokes theorems 
\beg
\label{stoke}
\int d^2 x \, \d_\mu f_\mu = - \oint dx_\mu \ep_{\mu\nu} f_\nu, ~~~~~  
\int d^2 x ~ \ep_{\mu\nu} \, \d_\mu f_\nu = \oint dx_\mu f_\mu  
\endeq
Consider for instance the contribution:
\beg
\label{3.20}
\int d^2 x ~ \langle \d_\mu \phi (x) \d_\mu \phi (0) \rangle 
= - \oint_\CC dx_\mu \, \ep_{\mu\nu} \, 
 \langle \phi (x) \d_\nu \phi (0) \rangle
\endeq
where the contour $\CC$ is the boundary of the disk.  Using
Eq. (\ref{corr2}) one finds\footnote{The 
 integral over $d\zbar$ originally has opposite sign due to 
$\ep_{\zbar z} = - \ep_{z\zbar}$, but for the contour $\CC$, 
the sense of integration over
$\zbar$ is reversed in comparison to $z$;  the integrals in Eq. 
(\ref{3.21}) are both usual Cauchy integrals.}  
\beg
\label{3.21}
\int d^2 x ~ \langle \d_\mu \phi (x) \d_\mu \phi (0) \rangle 
= 2\pi \( \oint  \frac{dz}{2\pi i} \inv{z} 
+ \oint \frac{d\zbar}{2\pi i} \inv{\zbar} \) = 4\pi 
\endeq
Similar reasoning gives 
\beg
\label{3.22} 
\int d^2 x ~ \langle \d_\mu \phid (x) \d_\mu \phid (0) \rangle = 4\pi , 
~~~~~
\int d^2 x ~ \langle \d_\mu \phid (x) \d_\mu \phi (0) \rangle = 0 
\endeq
We also need
\beg
\label{3.23}
\int d^2 x ~ \ep_{\mu\nu} \, 
\langle \d_\mu \phi (x) \d_\nu \phid (0) \rangle
= \oint_\CC dx_\mu \langle \phi(x) \d_\mu \phid (0) \rangle = 4\pi i 
\endeq
The same result holds with $\phi $ and $\phid$ interchanged.  Putting
this all together we obtain
\begin{eqnarray}
\label{3.24}
\sigma'_{xx} &=& \inv{2\pi g^2}  
 \( 1 + \(  \theta / \pi g^2  \)^2 \) 
\\ \nonumber
\sigma'_{xy} &=& \frac{i}{2\pi}  \frac{2\theta}{\pi g^4}   
\end{eqnarray}
The overall $1/2\pi$ is expected from the fact that $\sigma_{\mu\nu}$ 
has units of $e^2/h = 1/2\pi$. 

That $\sigma'_{xy}$ is imaginary for real $\theta$ can be traced
back to the ``$i$'' in the coupling to the gauge field in 
Eq. (\ref{2.13}).  
We now give arguments supporting the idea that for the computation
done in this section, we should analytically continue from euclidean
to Minkowski space.    
The reason 
has to do with the difference between the hermiticity properties
in first quantization verses second.  In first quantization, 
the hamiltonian $H$ is a hermitian operator $H^\dagger = H$.  
For our particular hamiltonian (\ref{2.5}) this follows from
$(-i\d_\mu)^\dagger = -i\d_\mu$ and $A^\dagger_{x,y} = A_{x,y}$.  However,
when working with the action $S$ as a conformal field theory,
 the natural reality properties of
the functional integral are different from this.  
Note for instance that the bosonization (\ref{2.8}) is at odds with the 
hermiticity properties of $\psi_{1,2}$.  Let us focus on the current
$j_\mu$ in Eq. (\ref{2.9}).  In the first quantized viewpoint,
$(\psi_{1,2})^\dagger = \psi_{1,2}^\dagger$ implies 
$(j_z )^\dagger = j_\zbar$, which requires $(\d_z \phi )^\dagger 
= \d_\zbar \phi$.  The latter is not the usual hermiticity
assumed in conformal field theory
 which is tied to the functional integral.  In
conformal field theory
 one considers rather $(\d_z \phi)^\dagger = \d_z \phi$.  
This suggests that to perform the computations in this section
meaningfully, one should analytically continue to Minkowski space.
The complex analytic structure of the conductivity tensor as
displayed in Eq. (\ref{3.2}) also allows the interpretation of
the analytic continuation of $\sigma_{xy}$ as a continuation from
euclidean to Minkowski space for the coordinates $x,y$. 
From the point of view of first quantization, this seems to
correspond having an imaginary vector potential.  
We point out that 
the importance of imaginary vector potentials 
for delocalization transitions has recently been recognized in
both electronic and biological systems \cite{Nelson}\cite{imag}\cite{Nelson2}.
For more discussion on this rather delicate point, see \cite{andre}.

We define now the critical condition
\beg
\label{3.27}
g^2 /2 = \pm \theta/ 2\pi
\endeq
There are two interpretations of this condition depending on
the prescription taken to make $\sigma_{xy}$ real, as we now describe. 

We can perform the analytic continuation to Minkowski space by
letting $\theta \to -i \theta$.  The critical condition then leads to
$\sigma' = \sigma^c$ with 
\beg
\label{E2}
\sigma^c_{xx} = 0 , ~~~~~ \sigma^c_{xy} = 1/\theta 
\endeq
Note that this is precisely the identification we made in arriving
at the zero mode constraint Eq. (\ref{2.18}).
In terms of $\tau$, the critical condition leads to 
\beg
\label{tauc}
\tau^c = \frac{\theta}{2\pi} \( 1 \pm  i \)
\endeq
For our original fermion model with $g=1$, the critical condition 
gives $\theta = \pm \pi$, so that $\tau^c = (1+i)/2 , (-1 + i)/2$.  
These are known to be the non-trivial fixed points of the 
$SL(2, \Zmath)$, one being related to the other by $\CT$.  Our
interpretation of the conformal field theory with the critical
condition is that it represents a pure, ideal  system consisting only of
the extended states, which are the extension of the edge
states into the bulk.  This pure system is known to have the same
conductance properties as a physical plateau in the presence of impurities
\cite{Prange}, so Eq. (\ref{E2}) leads us to this interpretation.   
We emphasize that we have not added any impurities yet, so that even
though Eq. (\ref{E2}) has the same properties as a plateau, there
are no real plateaux yet in our model since there are no localized 
states.  We cannot speak of the critical values of $\sigma_{xx}, 
\sigma_{xy}$ at a transition until impurities are added.  

The other possibility for making $\sigma_{xy}$ real is to analytically
continue $y \to i y$, keeping $\theta$ real as before, under which
$\sigma_{xy} \to \theta / \pi^2 g^4$, and $\sigma_{xx}$  remains 
unchanged.   Here the critical condition still leads to $\tau^c$,
and
\beg
\label{E4}
\sigma_{xy}^c = \pm \sigma_{xx}^c = 1/\theta
\endeq
For this prescription of analytic continuation, since
$\sigma_{xx}^c \neq 0$, the interpretation would be that we are
building into the model some features of the transition {\it with}
impurities, i.e. the property that $\sigma_{xx}^c \neq 0$.   
Since we will next add impurities as a perturbation, the 
previous interpretation is preferable.

The $SL(2, \Zmath)$ symmetry (\ref{modular}) does not have a simple
action on the conductivities since the latter are not a modular
transformation of $\tau$.  We can however point out the following.  
We are mainly interested in couplings satisfying the critical 
condition (\ref{3.27}).  Let us choose the positive sign in (\ref{tauc}),
and  construct a modular parameter out of $\sigma_{xx}, \sigma_{xy}$:  
\beg
\label{sigmod}
\varsigma = 2\pi i \sigma_{z\zbar} 
\endeq
Then one has:
\beg
\label{ST}
\varsigma = \CS \CT^n (\tau^c), ~~~~~~~~{\rm for} ~ \theta = -2\pi n
\endeq
So, for this restriction of the parameters, modular transformations
of $\tau$ induce modular transformations of $\varsigma$, and this
should lead to certain features of the phase diagram.

\bigskip\bigskip

{\color{\DG} 
\section{Adding a Circular Defect of Impurities} 
}

{\color{\DV}
\subsection{ The Impurity Potential}
}

We now introduce a potential $V$ representing some impurities in
the system.  
For the reasons explained in the Introduction we do not chose 
a random potential.  
We will take the potential
to be of  the form
\beg
\label{4.0}
V(x,y) =  V_0 \delta(r-r_0) 
\endeq
where $r^2 = x^2 + y^2$ and $r_0$ is some arbitrary radius.  This
corresponds to a circular defect line of impurities at $r=r_0$.  
As before, the sample is a disk of radius $r\to \infty$.  
This potential certainly incorporates a fixed realization of
impurities. As we will see, 
 the important advantage of our choice of potential is
that the  critical exponents do not depend on $V_0$.

The potential (\ref{4.0}) leads to a term in the action 
Eq. (\ref{2.7}) 
\begin{eqnarray}
\label{4.1} 
S_V &=& -i V_0 \int \frac{d^2x}{2\pi} ~ \delta (r-r_0) ~ \CO_V (x,y) 
\\ \nonumber
\CO_V &=& \psi_1^\dagger \psi_1 + \psi_2^\dagger \psi_2 
\end{eqnarray}
After bosonization, the model we will study has the action 
Eq. (\ref{2.16}) with the additional perturbation $S_V$, where 
$\CO_V = \cos \phi$.

\def\vphi{\varphi}

\bigskip

{\color{\DV}
\subsection{ The Defect Theory can be Mapped into 
 a Boundary Field Theory}
}

Let us map the theory to the euclidean cylinder as in section II, where the 
coordinates on the cylinder $t,\sigma$ are defined in Eq. (\ref{zcoor}). 
Since $r^2 = 
x^2 + y^2 = 2 z\zbar = 2e^{2t}$, the  defect is now a circle
along the circumference of the cylinder (the $\sigma$-direction) at
$t_0$ satisfying $r_0 = \sqrt{2} e^{t_0} $.  Without loss of generality
we let $t_0 = 0$.  

The theory can be solved by folding it onto a boundary field theory.
This folding procedure has been previously applied to other physical
problems\cite{Affleck}\cite{Saleur}\cite{Konik}\cite{AndAnd}.  
We first set $V_0 = 0$ and fold the conformal field theory.  In the
defect version of the problem any field can be separated into its pieces
on either side of the defect:
\beg
\label{4.2} 
\phi (t,\sigma) = \theta (t) \phi^{(+)} (t,\sigma) 
+ \theta (-t) \phi^{(-)} (t,\sigma) 
\endeq
where $\theta(t)$ is the step function,  $\theta(t) = 1$ for $ t>0$,
zero otherwise. 
Let $\phi_{L,R}^{(\pm)}$ denote the $L-R$ components of the scalar
field on either side of the defect: 
\beg
\label{plus}
\phi^{(\pm)} = \phi_L^{(\pm)} (t+ i\sigma) + \phi_R^{(\pm)} (t-i\sigma) 
\endeq
From these we define the following fields for $t>0$ only: 
\begin{eqnarray}
\label{4.3} 
\vphi &=& \vphi_L + \vphi_R, ~~~~~~~~~~ 
\vphi_L (t,\sigma) = \phi_L^{(+)} (t,\sigma)
, ~~~~\vphi_R (t,\sigma) = \phi_L^{(-)} (-t,\sigma) 
\\ \nonumber
\vphi' &=& \vphi'_L + \vphi'_R, ~~~~~~~~~~ 
\vphi'_L (t,\sigma) = \phi_R^{(-)} (-t,\sigma)
, ~~~~\vphi'_R (t,\sigma) = \phi_R^{(+)} (t,\sigma) 
\end{eqnarray}
As defined, the fields $\vphi_L , \vphi'_L$ $(\vphi_R , \vphi'_R)$
are functions of $t+i\sigma$ $(t-i\sigma)$, hence their L/R designations. 
Since they are defined only for $t>0$, they are fields in a theory
with a boundary at $t=0$.  Define now the even/odd combinations:
\beg
\label{4.4}
\vphi^{(e)} = \vphi + \vphi', ~~~~~ \vphi^{(o)} = \vphi - \vphi'
\endeq
The action for the boundary field theory
corresponding to the free hamiltonian can be written as 
\beg
\label{4.5} 
S_{\rm free} = \inv{2} \int_0^\infty dt \int d\sigma \inv{8\pi}  
\( (\d_\mu \vphi^{(e)} )^2 + (\d_\mu \vphi^{(o)} )^2 \) 
\endeq

The theory is not fully defined until the boundary conditions in
the conformal field theory are specified.  General properties  
of boundary conformal field theory were studied by Cardy\cite{Cardy2}. 
We first set the coupling to the gauge field to zero.  In the defect
description,  when $V=0$ 
the appropriate boundary condition is $\d_t \phi = 0$
since this corresponds to $\phi$ being continuous across the defect. 
Since $\phi_L , \phi_R$ are analytic functions of $w=t+i\sigma$,
$\wbar= t-i\sigma$ this implies that on the defect $\phi_L + \phi_R$
is a constant, which we take to be zero.  Letting 
$\phi = (\phi^{(+)} + \phi^{(-)})/2$ on the defect, in terms of
the boundary fields $\phi_L + \phi_R = 0$ reads 
\beg
\label{4.5b}
\phi_L^{(+)} + \phi_L^{(-)} + \phi_R^{(+)} + \phi_R^{(-)} = 
\vphi^{(e)} = 0, ~~~~~~~{\rm at} ~ t=0
\endeq

\def\vphid{\tilde{\vphi}}

Consider next the impurity operator $\CO_V$.  In the defect formulation
it is appropriate to take $\cos \phi$ at the defect to be the average
of its values on either side of the defect: 
\beg
\label{4.5c}
\cos \phi = \inv{2} \( \cos \phi^{(+)} + \cos \phi^{(-)} \) 
= \cos \[ (\phi^{(+)} + \phi^{(-)})/2 \]  
 \cos \[  (\phi^{(+)} - \phi^{(-)})/2 \]
,~~~~(t=0)
\endeq
Translating this to the boundary description, on the
boundary one has
\beg
\label{4.6}
\phi^{(+)} - \phi^{(-)} = \phi_L^{(+)} + \phi_R^{(+)}   
- \phi_L^{(-)} - \phi_R^{(-)}
= \vphi_L - \vphi_R - \vphi'_L + \vphi'_R = \vphid^{(o)} 
\endeq
where as before the dual field $\vphid^{(o)}$ is defined
as $\d_\mu \vphid^{(o)} = -i \ep_{\mu\nu} \d_\nu \vphi^{(o)}$.
  Thus on the boundary one has
\beg
\label{ov}
\CO_V  = \cos (\vphid^{(o)}/2) , ~~~~~~~~~~~(t=0) 
\endeq

\def\gh{\hat{g}}
\def\vphih{\hat{\vphi}}

Since the field $\vphi^{(e)}$ decouples from the boundary, we henceforth
drop it.  In order to match the normalization of the previous sections
we scale out the factor of $1/2$ in Eq. (\ref{4.5}) by now defining 
\beg
\label{newp}
\vphi \equiv \vphi^{(o)} / \sqrt{2} 
\endeq
The gauge field can now be restored;  for $A_\mu = \d_\mu \chi$,
the $\chi$ terms can also be folded as for  the $\phi$ field.
We finally obtain the action for the boundary theory:
\beg
\label{4.7}
S = \int_0^\infty dt \int d\sigma 
\( 
\inv{8\pi} (\d_\mu \vphi)^2 - \frac{i}{2\pi\gh} \ep_{\mu\nu} \d_\nu \vphi
\d_\mu \chi + \inv{2\pi \gh^2 } ( \d_\mu \chi )^2 
\)
-i  V_0 \int \frac{d\sigma}{2\pi} ~ \cos \( \frac{\vphid(0,\sigma)}{\sqrt{2}}\)
\endeq
where in the above equation $\chi \equiv \chi^{(o)}$. 
The coupling $\gh$ is related to $g$ as follows
\beg
\label{4.8} 
\gh = \sqrt{2} g
\endeq

Finally we need to impose a zero mode constraint as in 
Eq. (\ref{2.18}).  The original charge $Q$ in the defect theory
can be expressed as $Q=\int d\sigma \d_\sigma \phi^{(+)} /2\pi g$.  
At $t=0$ this is $Q=\int d\sigma \d_\sigma (\vphi^{(e)} + \vphi^{(o)} )
/4\pi g$. Dropping $\vphi^{(e)}$ and using Eq. (\ref{newp}) 
one has $Q=\int d\sigma \d_\sigma \vphi /2\pi \gh $.  Thus we impose
\beg
\label{4.9}
\frac{\theta}{2\pi \gh } \oint dx_\mu \d_\mu \vphi = \oint dx_\mu \d_\mu \chi
\endeq
where $\theta$ is the same as in the unfolded theory.  
(Here $x_\mu $ denotes $t,\sigma$.) 
The couplings $g, \theta$ are real, which
leads to real anomalous dimensions of operators, and the critical 
condition is (\ref{3.27}).     

The original fermion model is characterized by the values
$\gh = \sqrt{2}, ~~ \theta = \pi$. 
When $\theta = 0$, $\gh = \sqrt{2}$ is the self-dual point of the 
$SL(2,\Zmath)$ symmetry described in section II, and $V_0$ corresponds
to a (boundary) marginal perturbation  of scaling dimension $1$.  

Our model still possesses the gauge invariance (\ref{gauge}) with
$g$ replaced by $\gh$.  A 2-boson model with topological coupling
and a similar boundary interaction was considered in \cite{Callan};
again the important difference
is the gauge invariance of our model.  

\def\wbar{{\bar{w}}}

The new feature of the boundary version in comparison to what we had
in section II  is the boundary condition at
$t=0$, which now depends on $\theta$.  Letting $\vphi \to \vphi + 
\delta \vphi$ and  requiring $\delta S = 0$ on the boundary leads
to the equation of motion (when $V=0$):
\beg
\label{4.10b} 
-\inv{4\pi} \d_t \vphi + \frac{i}{2\pi \gh} \d_\sigma \chi = 0 , 
~~~~~(t=0)
\endeq
Imposing the zero mode constraint Eq. (\ref{4.9}) one finds 
\beg
\label{4.13}
\( \gh^2 /2  + \theta/2\pi  \) \d_w \vphi_L = 
- \( \gh^2 /2  - \theta/2\pi  \) \d_\wbar \vphi_R 
~~~~~~(t=0)
\endeq

This is an interesting relation since it links the modular properties
of the couplings $g,\theta$ 
with the spacial coordinates.
  Namely, going to Minkowski space and defining
the light-cone coordinates $v= t+\sigma$, $\bar{v} = t- \sigma$,
the boundary condition reads
\beg
\label{4.12}
\( \bar{\tau} \d_v - \tau \d_{\bar{v}} \) \vphi = 0 
\endeq
where $\tau$ is the modular parameter (\ref{modular}) with
$g\to \gh$ and $\bar{\tau} = \tau^* $.

We will need  the anomalous scaling dimension of the boundary
operator $\CO_V$.   We first gauge away $\chi$;  as before the 
zero mode constraint leads to the transformation (\ref{2.28}) 
with $g\to \gh$ so that 
\beg
\label{4.14}
\CO_V \to \cos 
\( \[  \( 1 + {\theta}/{\pi \gh^2 } \) \vphi_L 
 -   \( 1 - {\theta}/{\pi \gh^2 } \) \vphi_R \]
/\sqrt{2}  
\) 
\endeq

Viewing $\sigma $ as the ``time'' the hamiltonian can be written as 
\beg
\label{ham}
H = \( \inv{8\pi} \int_0^\infty  dt ~ (\d_w \vphi_L )^2 + 
(\d_\wbar \phi_R)^2 \) + \frac{i V_0}{2\pi} \, \CO_V (t=0) 
\endeq
The theory can now be rewritten using only a left-moving field. 
Using arguments found in \cite{Cardy2}, the boundary condition
Eq. (\ref{4.13}) allows us to view $\vphi_R$ as an analytic continuation
of $\vphi_L$.  Using Eq. (\ref{4.13}) we make the substitution  
\beg
\label{extend}
 \vphi_R 
= - \frac{( {\gh^2}/{2} + \theta/2\pi  )}{(\gh^2 /2 - \theta/2\pi )} ~ \vphi_L  
\endeq
 into Eq. (\ref{ham}).   Defining a rescaled field
\beg
\label{hatphi}
\vphih_L =  
\frac{ \sqrt{ 1 + \( \theta/\pi \gh^2 \)^2 }}{\( 1 - \theta/\pi \gh^2 \) } 
~ \vphi_L 
\endeq
one finds 
\beg
\label{ham2}
H = \( \inv{4\pi} \int d\sigma (\d_w \vphih_L )^2 \) 
+ \frac{i V_0}{2\pi} \cos \( \sqrt{2} a \vphih_L \) 
\endeq
where 
\beg
\label{aa}
a = \frac{ \( 1-\( \theta/\pi \gh^2 \)^2 \)}{
\sqrt{ 1 + \( \theta /\pi \gh^2 \)^2 } }
\endeq

Since the rescaled field has the two-point function 
$\langle \vphih_L (z) \vphih_L (0) \rangle = - \log z$, 
and 
\beg
\label{vev}
\langle e^{i\sqrt{2} a \vphih_L (z) } e^{-i \sqrt{2} a \vphih_L (0)} 
\rangle = 1/{z^{2a^2}}~, 
\endeq
the anomalous scaling dimension is 
\beg
\label{4.27}
[\CO_V]  \equiv {\rm dim} ( \CO_V) = a^2 
\endeq

\bigskip

{\color{\DV}
\subsection{ Critical Exponents}
}

The model of the last section has a single energy scale $M$ with units
of mass set by $V_0$. Since the action is dimensionless, the
dimension of $V_0$ is $1-[\CO_V]$.  The energy scale $M$ defines
a correlation length $M = 1/\xi_c$, thus 
\beg
\label{5.1} 
\xi_c \propto V_0^{-1/(1-[\CO_V])}
\endeq

Let $\rho (E)$ denote the density of states, so that $\rho (E) dE$
represents a number  of states per unit volume.  It can be expressed as
\beg
\label{5.2}
\rho (E)  = \inv{\CV} {\rm Tr} ~ \delta (H-E) 
= \inv{\pi \CV } \lim_{\eta \to 0^+} {\rm Im} ~ {\rm Tr} \( 
\inv{H - E  - i \eta} \) 
\endeq
where $\CV$ is the 2d volume.  This in turn can be expressed as a 
retarded Green function 
\beg
\label{5.4}
\rho (E)  = \inv{\pi} \lim_{\eta \to 0^+} {\rm Im} 
\Langle \psi^\dagger_{\vep } (x) \psi_{\vep} (x) \Rangle
~~~~~ \vep = E  + i\eta
\endeq
In accordance with the discussion in section III, we define a critical
density of states $\rho^c$ from the above formula with $\vep = 0$.  
As we argued in section III 
for $\sigma'_{\mu\nu}$, this quantity should represent the density
of states near the critical point.  Since both fields in Eq. (\ref{5.4})
are at the same point $x$, $\rho^c$ is a one-point function of the 
operator $\psi^\dagger \psi = \psi_1^\dagger \psi_1 + \psi_2^\dagger 
\psi_2$.  From Eq. (\ref{4.1}) this operator is the operator $\CO_V$:
\beg
\label{e1}
\rho^c \propto \langle \CO_V \rangle
\endeq
From the scaling dimension of $\langle \CO_V \rangle$ we know that
\beg
\label{e2}
\langle \CO_V \rangle \propto \( \xi_c \)^{-[\CO_V]} 
\endeq

Next, recall that the number of states per unit volume in a Landau level
is $B/2\pi$.  Thus, $\rho^c$ should  scale with $B$, so that 
$\rho^c \propto |B-B_c|$.  Using Eqs. (\ref{e1})(\ref{e2}) one obtains the
relation (\ref{1.2}) with    
\beg
\label{expon}
\nu = 1/[\CO_V]
\endeq
Alternatively we can argue that since since $\rho (E) dE$ has units
of inverse volume, in two dimensions $\rho$ has dimensions of energy.
Letting $\rho^c \propto E$,  one then  finds 
\beg
\label{ecrit}
\xi_c \propto |E-E_c|^{-1/[\CO_V]}
\endeq
where here $E_c = 0$.  
The connection between the two equations (\ref{ecrit}) and 
(\ref{1.2}) is due to the fact that the Fermi energy is 
proportional $B$ in a system with Landau levels.

In order to satisfy the critical condition  $\sigma'_{xx} = 0$,
from Eq. (\ref{3.27}) we impose $\theta / \pi g^2 = \pm 1$.  Using 
Eq. (\ref{4.14}) one has  $\theta/\pi \gh^2 = \pm 1/2$.  Inserting
this into Eq. (\ref{4.27}) we obtain the value $\nu = 20/9$ quoted
in the Introduction.  Note that this exponent only depends on
the ratio $\theta/ g^2$.   
This value of $\nu$ holds for {\it any} $g,\theta$ with 
$\sigma'_{xx} = 0$.  In particular $\sigma'_{xy} = 1/\theta$ is left
unconstrained, and this indicates that the exponent $\nu$ is universal.

\bigskip

{\color{\DG}
\section{ 
S-matrices designed from the modular parameter $\tau$ 
exhibit a series of plateaux}
}

In this section we propose an intriguing connection with the
so-called staircase model, whose interesting  properties were 
first witnessed  by Al. Zamolodchikov\cite{Alyosha},  and were considered
mysterious at the time.

The conformal model of section II is essentially
that of a single massless scalar field with coupling constants
defining a modular parameter $\tau$ and possessing an $SL(2,\Zmath)$
symmetry.  
The original staircase model was characterized by a bulk S-matrix
for a single  massive particle.  Though it is the boundary version
that is more appropriate to our problem, let us begin by showing
how the structure of the S-matrix follows quite naturally from what
we have done.     
The energy and momentum of a relativistic particle 
can be parameterized by a rapidity $\beta$:
\beg
\label{6.1} 
E = m \cosh \beta   , 
~~~~~~
P = m\sinh \beta
\endeq
where $m$ is the mass of the particle.  
The bulk S-matrix describing the 2-particle to 2-particle
scattering must satisfy crossing symmetry and unitarity\cite{ZamoZamo}: 
\beg
\label{6.2}
S(\beta ) = S(i\pi - \beta) , ~~~~~  S(\beta) S(-\beta ) = 1
\endeq
S-matrices satisfying the above functional equations are usually
built out of products of the minimal factors: 
\beg
\label{6.3} 
S (\beta) =  \frac{  \sinh \beta - i \sin \pi \gamma }{\sinh \beta 
+ i \sin \pi \gamma} 
\endeq
where $\gamma$ is a parameter related to the coupling constants 
of the theory.  
Let us attempt to relate the S-matrix  to the physics
of our problem by using the $SL(2, \Zmath)$ symmetry to relate
$\gamma$ to $\tau$.  
The S-matrix has the following symmetries:
\beg
\label{6.4} 
\gamma \to \gamma + 2, ~~~~\gamma \to 1 - \gamma
\endeq
It does not seem possible  for the S-matrix
to possess the full $SL(2,\Zmath)$ symmetry.  
The first symmetry in the above formula suggests the identification
$\gamma = \tau$, since it would correspond to the transformation 
$\CT^2$, where $\CT$ is defined in Eq. (\ref{Z7}).  
Next consider the transformation $\gamma \to 1-\gamma$.  The modular
parameters $\tau$ satisfying the critical condition Eq. (\ref{3.27}) 
come in complex conjugate pairs Eq. (\ref{tauc}). 
It is natural then that the S-matrix not distinguish between these two
critical $\tau$'s.  Requiring then that $\gamma \to 1-\gamma$ is
equivalent to $\tau_+^c \to \tau_-^c$ requires $g=1$, or equivalently
$\theta = \pi$,  in which case $\gamma = \tau_\pm^c = (1 \pm i)/2$.   
Finally we perturb away from the critical  condition by deforming
$g$ away from $1$ in $\tau$, but keeping $\theta = \pi$.  Because
of the form of the critical condition, we can  
just as well view this deformation  as a deformation of $\theta$
away from $\pi$.  Thus we  write this as $g^2/2 \to \theta_0/2\pi$,  
and identify $\gamma$ as follows:
\beg
\label{6.7}
\gamma = \tau_\pm = \tau(\theta = \pi, g^2/2 = \pm \theta_0 /2\pi) 
= \inv{2} \pm i \frac{\theta_0}{2\pi}  
\endeq
The S-matrix is then 
\beg
\label{6.8} 
S (\beta) =  \frac{  \sinh \beta - i \cosh  \theta_0/2  }{\sinh \beta 
+ i \cosh \theta_0/2 } 
\endeq
   
It was discovered by Al. Zamolodchikov that the above S-matrix leads
to a free energy with some remarkable properties.   The model
can be studied on a cylinder of length $l$ and radius $R$.
  Imposing periodic boundary conditions in
the $l$-direction, the free energy $\CE (R)$ was computed starting
from the S-matrix by means  of the thermodynamic Bethe ansatz\cite{Yang}.
It was found in \cite{Alyosha} that as one varies $R$ 
the free energy 
$\CE (R)$ goes through a series of plateaux, i.e. the free-energy
as a function of $R$ resembles a staircase.

\def\bh{\hat{\beta}}

Let us try and be more specific now about the relation with our model.  
The boundary theory we obtained is defined by the action 
(\ref{4.7}).   Let us first decouple the gauge field, $\chi = 0$.  
Noting that $(\d_\mu \vphi )^2 = - (\d_\mu \vphid )^2$, letting
$\vphid \to i \phi$ we obtain from (\ref{4.7}): 
\beg
\label{shG}
S = \int_0^\infty dt \int d\sigma  
\[ \inv{8\pi}  (\d_\mu \phi )^2 + \Lambda \cosh (\bh \phi ) \] 
~ + \frac{V_0}{2\pi i}  \int d\sigma ~ \cosh (\bh \phi /2 ) 
\endeq
with $\bh = \sqrt{2}$.  Our model has zero bulk coupling $\Lambda = 0$.
We have included a $\Lambda$ term since as written the above action
defines the boundary sinh-Gordon model.  It is an integrable
model that can be studied using the framework developed in 
\cite{GZ}.  The bulk S-matrix is known to be of the form 
 Eq. (\ref{6.3}) with 
\beg
\label{gam}
\gamma = \frac{\bh^2}{2 + \bh^2} 
\endeq
  
First note that our model leads to $\bh = \sqrt{2}$ and 
$\gamma = 1/2$.  This is precisely the same as 
in Eq. (\ref{6.7}) when $\theta_0 = 0$.  
It is therefore clear  that our model 
(\ref{4.7}) is very closely related to the boundary
sinh-Gordon model with $\bh = \sqrt{2}$ with $\Lambda = 0$ 
and $\gamma$ analytically continued as in (\ref{6.7}).  
This analytic continuation of $\gamma$ incorporates 
the topological coupling to the gauge field $\d\chi$, and amounts to 
continuing the coupling $g^2/2$ to the modular parameter $\tau$.   
In terms of $\bh$, the above analytic continuation of $\gamma$
corresponds to a simple phase:
\beg
\label{phase}
\bh = \sqrt{2} e^{i\alpha} , ~~~~~
\cos 2 \alpha =  \frac{  1 - (\theta_0 /\pi )^2 }{1+ (\theta_0 /\pi )^2 }  
\endeq

A boundary version of the staircase model was studied in 
\cite{Lesage}.\footnote{This paper also includes a useful reproduction of
the bulk results in \cite{Alyosha}.}  Indeed it was found that
the boundary entropy reveals a series of plateaux as a function of
renormalization group scale.  

Keeping within the scope of this paper we refrain from going any more
deeply into the integrability structures that allow a detailed study
of this proposal, but will return to this in a future publication.


\vfill\eject

{\color{\DG}
\section{Discussion}
}

In summary, we have constructed a reasonably simple model which
we argued describes the critical properties of Quantum Hall plateau
to plateau transitions.  It is a $c=1$ conformal field theory
mainly characterized by a gauge symmetry and the  
$SL(2, \Zmath)$ symmetry which follows from it.
Adding impurities singles out
a particular operator $\CO_V$,  whose scaling dimension, as  
computed in the conformal field theory, follows largely from
the gauge invariance, and leads to the exponent $\nu = 20/9$.

Experimental errors at this time are perhaps too large to
distinguish between $\nu = 2.3$ and $\nu = 2.2$.  However if
the measurements continue to indicate values closer to
$\nu = 20/9$, as in \cite{Sondhi}, this suggests that  the transition
is in a different universality class than the percolative class 
of the network/Anderson models, assuming the numerical work
on the latter is correct.  Our result appears to be more
consistent with the model used in Ando's simulation\cite{Ando}, 
which gave $\nu = 2.2 \pm .1$.      

Since the gauge symmetry is an important feature of our model it
can perhaps be viewed as a simplified version of 
the Yang-Mills theories in higher dimensions which exhibit
the $SL(2, \Zmath)$ electric/magnetic duality\cite{Seiberg}.

 The boundary staircase model described in section V is
 a promising candidate for a model that exhibits a series of
plateaux transitions.    

\bigskip\bigskip

{\color{DarkGreen}
\section{ Acknowledgments}
} 
I would like to thank the Institute for Theoretical Physics in Santa Barbara,
the organizers of the program {\it Quantum Field Theory in Low 
Dimensions:  From Condensed Matter to Particle Physics} in the Spring of
1997,  and the many participants who helped interest me   
in this problem, especially Andreas Ludwig for many discussions.  
  This work is supported in part by
the National Young Investigator Program of the NSF.     

\bigskip

\section{Addendum}

After the first version of this paper appeared,  we reported a study of the 
relevance of disorder in \cite{andre}.   There we gave evidence for
two possible universality classes in the presence of disorder, in one
of which all disorder is driven irrelevent by the presence of disorder
in the gauge field.


\begin{references}
\bibitem{Huckestein} B. Huckestein, Rev. Mod. Phys. 67 (1995) 357.

\bibitem{Anderson0} P. W. Anderson,  Phys. Rev. 112 (1958) 1900. 

\bibitem{Anderson} E. Abrahams, P. W. Anderson, D. C. Licciardello
and T. V. Ramakrishnan, Phys. Rev. Lett. 42 (1979) 673.

\bibitem{Wegner}  F. J. Wegner, Z. Phys. B 35 (1979) 208; ibid B51 (1983) 
579.  

\bibitem{Pruisken} A. M. M. Pruisken, Nucl. Phys. B285[FS19] (1987) 719; 
Nucl. Phys. B290[FS20] (1987) 61. 

\bibitem{Prange}  {\it The Quantum Hall Effect}, 
eds.  R. E. Prange and S. M. Girvin, 
Springer-Verlag, Berlin, 1987. 

\bibitem{Zirnbauer}  M. Zirnbauer,  Ann. d. Physik 3 (1994) 513.

\bibitem{Haldane}  F. D. M. Haldane, J. Appl. Phys. 57 (1985) 3359.  


\bibitem{Chalker} J. T. Chalker and P. D. Coddington, J. Phys. C 21
(1988) 2665.   

\bibitem{Chalker2}  C.-M. Cho and J. T. Chalker, cond-mat/9605073.

\bibitem{Ludwig}  A. W. W. Ludwig, M. P. A. Fisher, R. Shankar
and G. Grinstein, Phys. Rev. B50 (1994) 7526.

\bibitem{Laughlin}  R. B. Laughlin,  Phys. Rev. B 23 (1981) 5632.

\bibitem{Halperin}  B. I. Halperin, Phys. Rev. B 25 (1982) 2185. 

\bibitem{andre}  A. LeClair, {\it On the Relevance of Disorder in
Quantum Hall Plateaux Transitions},  cond-mat/9905222.    

\bibitem{Fisher}  P. A. Lee and D. S. Fisher, Phys. Rev. Lett 47 (1981) 882.

\bibitem{BPZ}  A. A. Belavin, A. M. Polyakov and A. B. Zamolodchikov,
Nucl. Phys. B241 (1984) 333.


\bibitem{Witten}  K. S. Narain, M.H. Sarmadi and E. Witten, 
Nucl. Phys. B279 (1987) 369.

\bibitem{Wilczek}  A. Shapere and F. Wilczek, Nucl. Phys. B320 (1989) 669. 

\bibitem{Cardy1} J. L. Cardy, Nucl. Phys. B205[FS5] (1982) 17. 

\bibitem{Kiv1} S. Kivelson, D.-H. Lee and S.-C. Zhang, Phys. Rev. B 46 (1992)
2223.


\bibitem{dual}  C. A. L\"utken and G. G. Ross, Phys. Rev. B45 (1992) 11837;
ibid. 48 (1993) 2500;  
 C. A. L\"utken, Nucl. Phys. B396 (1993) 670; 
C. P. Burgess and C. A. L\"utken, ibid. B500 (1997)
367. 

\bibitem{Frad}  E. Fradkin and S. Kivelson, Nucl. Phys. B474 (1996) 543. 

\bibitem{dual4} N. Taniguchi, cond-mat/9810334.

\bibitem{dual5}  B. P. Dolan, cond-mat/9805171; ibid./9809294. 
 
\bibitem{Coleman}   S. Coleman, Phys. Rev. D11 (1975) 2088.

\bibitem{Ginsparg}  P. Ginsparg, Les Houches lectures 1988, eds. 
E. Br\'ezin and J. Zinn-Justin, Elsevier 1990.  

\bibitem{Mathieu}  P. Di Francesco, P. Mathieu and D. S\'en\'echal, 
{\it Conformal Field Theory}, 1997, Springer-Verlag, New York. 

\bibitem{Pruisken2}  A. M. M. Pruisken, Phys. Rev. Lett. 61 (1988) 1297.

\bibitem{fourseven}  G. V. Mil'nikov and I. M. Sokolov, JETP Lett 48 (1988) 
536. 

\bibitem{Wei}  H. P. Wei, L. W. Engel and  D. C. Tsui, 
Phys. Rev. B 50 (1994) 14609.  

\bibitem{Koch} S. Koch, R. J. Haug, K. von Klitzing and K. Ploog, 
Phys. Rev. Lett. 67 (1991) 883.  

\bibitem{Sondhi} D. Shahar, D. C. Tsui, M. Shayegan, E. Shimshoni
and S. L. Sondhi, Phys. Rev. Lett. 79 (1997) 479. 

\bibitem{Kivelson}  D.-H. Lee, Z. Wang and S. Kivelson,  Phys. Rev.
Lett 70 (1993) 4130. 

\bibitem{Ando}  T. Ando, J. Phys. Soc. Jpn. 61 (1992) 415. 

\bibitem{Schwarz} M. B. Green and J. H. Schwarz, Phys. Lett. 149B (1984) 117.

\bibitem{Mahan} G. D. Mahan, {\it Many-Particle Physics}, 1981, Plenum
Press, New York.

\bibitem{Thouless}  D. J. Thouless, Phys. Rep. 13 (1974) 93.

\bibitem{Stone}  A. J. McKane and M. Stone,  Ann. Phys. 131 (1981) 36. 

\bibitem{Nelson}  N. Hatano and D. R. Nelson, Phys. Rev. Lett. 77 (1996) 570;
Phys. Rev. B 56 (1997) 8651.

\bibitem{imag}  T. Morinari, cond-mat/9903295. 

\bibitem{Nelson2}  K. A. Dahmen, D. R. Nelson and
N. M. Shnerb, cond-mat/9903276.  

\bibitem{Wen}  X. G. Wen, Phys. Rev. B 41 (1990) 12838.

\bibitem{Affleck}  E. Wong and I. Affleck, Nucl. Phys. B417 (1994) 403. 

\bibitem{Saleur}  P. Fendley, A. W. W. Ludwig and H. Saleur, 
Phys. Rev. Lett. 74 (1995) 3005;  Phys. Rev. B52 (1995) 8934. 

\bibitem{Konik}  R. Konik and A. LeClair, Nucl. Phys. B538 (1999) 587. 

\bibitem{AndAnd}  A. LeClair and A. W. W. Ludwig, hep-th/9708135,
to appear in Nucl. Phys. B. 

\bibitem{Cardy2} J. L. Cardy, Nucl. Phys. B270 (1986) 186; ibid 275 (1986) 
200; ibid 324 (1989) 581. 

\bibitem{Callan}  C. G. Callan, I. R. Klebanov, J. M. 
Maldacena and A. Yegulalp, Nucl. Phys. B443 (1995) 444.  


\bibitem{Alyosha}  Al. B. Zamolodchikov, {\it Resonance Factorized
Scattering and Roaming Trajectories}, Ecole Normale Sup\'erieure
preprint, ENS-LPS-335, 1991, otherwise unpublished. 

\bibitem{ZamoZamo}  A. B. Zamolodchikov and Al. B. Zamolodchikov,
Ann. Phys. 120 (1979) 253. 

\bibitem{Yang} C. N. Yang and C. P. Yang, J. Math. Phys. 10 (1969) 1115. 

\bibitem{GZ}  S. Ghoshal and A. B. Zamolodchikov, Int. J. Mod. Phys.
A9 (1994) 3841.

\bibitem{Lesage}  F. Lesage, H. Saleur and P. Simonetti, 
Phys. Lett. B427 (1998) 85.

\bibitem{Seiberg} N. Seiberg and E. Witten, Nucl. Phys. B426 (1994) 19.




\end{references}
\end{document}